\documentclass[12pt]{article}

\begin{document}

\begin{titlepage}
 
\begin{center}
\vskip 1.0cm
NATIONAL ACADEMY OF SCIENCES OF UKRAINE
\vskip 0.4cm
INSTITUTE FOR NUCLEAR RESEARCH
\end{center}

\begin{flushright}
\vskip 1.5cm
Manuscript
\end{flushright}

\begin{center}
\vskip 2.0cm
\textbf{\large{Poda Denys Valentynovych}}\footnote{\noindent e-mail:~poda@kinr.kiev.ua~~A full version of the dissertation in Ukrainian can be found in: http://lpd.kinr.kiev.ua/poda/D.V.Poda-PhD-thesis.pdf}
\end{center}

\begin{flushright}
\vskip 1.0cm
UDK~539.165
\end{flushright}

\begin{center}
\vskip 1.0cm
\textbf{\Large{Double beta decay of $^{64,~70}$Zn and $^{180,~186}$W isotopes}}

\vskip 1.0cm
01.04.16 $-$ physics of nuclei, elementary particles, and high energies

\vskip 0.4cm
A thesis for the Ph.D. Degree (Physics and Mathematics)
\end{center}

\begin{flushright}
\vskip 2.0cm
\large{Thesis Advisor:}
\vskip 0.2cm
\large{Dr.Sci. F.A. Danevich}
\end{flushright}

\begin{center}
\vskip 3.0cm
Kyiv $-$ 2009
\end{center}

\end{titlepage}

\newpage
\clearpage

\begin{center}
\textbf{ABSTRACT}
\end{center}

\vskip 0.4cm
The results of the experimental investigations of double beta processes in Zinc and Tungsten isotopes with the 
help of middle volume (117 g, 168 g and 699 g) low-background ZnWO$_4$ crystal scintillators are presented. The 
experiment was carried out in the low-background ``DAMA/R\&D'' set-up at the Gran Sasso National Laboratories of 
the INFN (Italy) at a depth of $\approx$3600 m w.e. The total measurement time exceeds ten thousand hours. New improved half-life limits on double electron capture and electron capture with positron emission in $^{64}$Zn have been set: $T^{2\nu 2K}_{1/2}$($^{64}$Zn) $\geq$ 6.2(6.3) $\times$ 10$^{18}$ yr, $T^{0\nu 2\varepsilon}_{1/2}$ ($^{64}$Zn) $\geq$ 1.1(2.8) $\times$ 10$^{20}$ yr, $T^{2\nu \varepsilon \beta^+}_{1/2}$($^{64}$Zn) $\geq$ 0.7(2.1) $\times$ 10$^{21}$ yr, and $T^{0\nu \varepsilon \beta^+}_{1/2}$($^{64}$Zn) $\geq$ 4.3(5.7) $\times$ 10$^{20}$ yr, all the limits are at 90\% (68\%) C.L. The positive indication on the $\varepsilon \beta^+$ decay of $^{64}$Zn with $T^{(2\nu + 0\nu) \varepsilon \beta^+}_{1/2}$ ($^{64}$Zn) = (1.1 $\pm$ 0.9) $\times$ 10$^{19}$ yr suggested in [Appl. Radiat. Isot. 46 (1995) 455--456] is fully discarded by the present experiment. To date only two nuclei ($^{40}$Ca and $^{78}$Kr) among 34 potentially ``2$\beta^+$ active'' nuclides were studied at the similar level of sensitivity. However, it is worth noting that the theoretical predictions are still higher.

The half-life limits on the 2$\beta$ processes in $^{70}$Zn, $^{180}$W, and two neutrino mode of 2$\beta^-$ decay in $^{186}$W established in the present work on the level of 10$^{17}-10^{20}$~yr are one order of magnitude higher than those set in previous experiments.

Energy resolution, relative light output, $\alpha/\beta$ ratio, decay time, pulse-shape discrimination between $\alpha$  particles and $\gamma$ rays ($\beta$ particles), and radioactive contamination of CdWO$_4$, PbWO$_4$ (undoped, and doped by F, Eu, Mo, Gd, S), and ZnWO$_4$ crystal scintillators were studied. Pulse-shape discrimination ability of PbWO$_4$ and ZnWO$_4$ crystal scintillators were realized for the first time. The first result of low-background measurement with small volume ZnWO$_4$ (mass of 4.5 g) gave reasons for extensive research work in the Institute for Scintillation Materials (Kharkiv, Ukraine) in order to optimize the growth conditions with the aim of producing high quality large-volume ZnWO$_4$ crystal scintillators. Applicability of these scintillators to search for double beta decay was proved.

The time-amplitude analysis, the pulse-shape discrimination, and the Monte Carlo simulation were applied in addition to the ICP-MS measurements to estimate radioactive contamination of the ZnWO$_4$ detectors. We have found ZnWO$_4$ crystal scintillators extremely radiopure detectors with typical contamination at the level of $\mu$Bq/kg ($^{228}$Th and $^{226}$Ra), $\leq$ 0.06 mBq/kg ($^{210}$Po), total $\alpha$ activity (U/Th) $0.2-0.4$ mBq/kg, $\leq$ 0.4 mBq/kg ($^{40}$K), $\leq$ 0.05 mBq/kg ($^{137}$Cs), $\leq$ 0.4 mBq/kg ($^{90}$Sr-$^{90}$Y), $\leq$ 0.01  mBq/kg ($^{147}$Sm), and $\leq$ 3 mBq/kg ($^{87}$Rb). 
Our investigations with zinc tungstate crystals have demonstrated a good potential of ZnWO$_4$ scintillators for the next generation double beta decay and cryogenic dark matter experiments, in particular for EURECA project, where a multi-element target with the total mass up to 1 t is planned for confirming dark matter signal. High abundance of $^{64}$Zn (48.3\%) allows to build a large scale double beta experiment without expensive isotopic enrichment. An experiment involving $\approx$10 tons of ZnWO$_4$ crystals (9 $\times$ 10$^{27}$ nuclei of $^{64}$Zn) could reach the half-life sensitivity up to 3 $\times$ 10$^{28}$ yr (supposing zero background during ten years of measurements). Such a sensitivity could contribute to our understanding of the neutrino mass mechanism and right-handed currents in neutrinoless processes. The two neutrino double electron capture should be surely observed: in accordance with the theoretical expectations $T_{1/2}$ for $2\nu 2\varepsilon$ process is predicted on the level of 10$^{25}-10^{26}$ yr.

A new project of high sensitive $2\beta^-$  experiment was proposed. For this purpose, the new detection system with high light collection and energy resolution was developed, and PbWO$_4$ crystals were also discussed as high-efficiency 4$\pi$ active shield and light guides in $^{116}$Cd double beta decay experiment with enriched $^{116}$CdWO$_4$ crystal scintillators. The sensitivity of such an experiment (in terms of the half-life limit) is estimated as $lim~T^{0\nu 2\beta^-}_{1/2}$($^{116}$Cd) $\approx$ 10$^{26}$~yr, which corresponds to the effective Majorana neutrino mass $\left\langle m_{\nu} \right\rangle \approx 0.07$ eV.


\vskip 2.0cm
\begin{center}
\textbf{PUBLICATIONS}
\end{center}

\vskip 0.2cm
\noindent [1] P. Belli, R. Bernabei, F. Cappella, R. Cerulli, C.J. Dai, F.A. Danevich, B.V. Grinyov, 
A. Incicchitti, V.V. Kobychev, V.M. Mokina, L.L. Nagornaya, S.S. Nagorny, S. Nisi, F. Nozzoli, D.V. Poda, 
D. Prosperi, V.I. Tretyak, S.S. Yurchenko, 
\textit{``Searches for $2\beta$ decay of Zinc and Tungsten with the help of low-background ZnWO$_4$ crystal scintillators''}, 
Nucl. Phys. B 826 (2009) 256${-}$273. 

\vskip 0.4cm
\noindent [2] P. Belli, R. Bernabei, F. Cappella, R. Cerulli, C.J. Dai, F.A. Danevich, B.V. Grinyov, 
A. Incicchitti, V.V. Kobychev, L.L. Nagornaya, S.S. Nagorny, F. Nozzoli, D.V. Poda, D. Prosperi, 
V.I. Tretyak, S.S. Yurchenko, 
\textit{``Search for $2\beta$ processes in $^{64}$Zn with the help of ZnWO$_4$ crystal scintillator''}, 
Phys. Lett. B 658 (2008) 193${-}$197.

\vskip 0.4cm
\noindent [3] L. Bardelli, M. Bini, P.G. Bizzeti, F.A. Danevich, T.F. Fazzini, N. Krutyak, V.V. Kobychev, 
P.R. Maurenzig, V.M. Mokina, S.S. Nagorny, M. Pashkovskii, D.V. Poda, V.I. Tretyak, S.S. Yurchenko, 
\textit{``Pulse shape discrimination with PbWO$_4$ crystal scintillators''},  
Nucl. Instr. Meth. A 584 (2008) 129${-}$134.

\vskip 0.4cm
\noindent [4] L.L. Nagornaya, A.M. Dubovik, Yu.Ya. Vostretsov, B.V. Grinyov, F.A. Danevich, K.A. Katrunov, 
V.M. Mokina, G.M. Onishchenko, D.V. Poda, N.G. Starzhinskiy, I.A. Tupitsyna, 
\textit{``Growth of ZnWO$_4$ crystal scintillators for high sensitivity $2\beta$ experiment''}, 
IEEE Trans. Nucl. Sci. 55 (2008) 1469${-}$1472.

\vskip 0.4cm
\noindent [5] P. Belli, R. Bernabei, N. Bukilic, F. Cappella, R. Cerulli, C.J. Dai, F.A. Danevich, 
J.R. de Laeter, A. Incicchitti, V.V. Kobychev, S.S. Nagorny, S. Nisi, F. Nozzoli, D.V. Poda, D. Prosperi, 
V.I. Tretyak, S.S. Yurchenko, 
\textit{``Investigation of $\beta$ decay of $^{113}$Cd''}, 
Phys. Rev. C 76 (2007) 064603, 10 p.

\vskip 0.4cm
\noindent [6] F.A. Danevich, A.Sh. Georgadze, V.V. Kobychev, B.N. Kropivyansky, S.S. Nagorny, A.S. Nikolaiko, 
D.V. Poda, V.I. Tretyak, I.M. Vyshnevskyi, S.S. Yurchenko, B.V. Grinyov, L.L. Nagornaya, E.N. Pirogov, 
V.D. Ryzhikov, V.B. Brudanin, Ts. Vylov, A. Fedorov, M. Korzhik, A. Lobko, O. Missevitch, 
\textit{``Application of PbWO$_4$ crystal scintillators in experiment to search for $2\beta$ decay of $^{116}$Cd''}, 
Nucl. Instr. Meth. A 556 (2006) 259${-}$265.
\newpage
\vskip 0.4cm
\noindent [7] L. Bardelli, M. Bini, P.G. Bizzeti, L. Carraresi, F.A. Danevich, T.F. Fazzini, B.V. Grinyov, 
N.V. Ivannikova, V.V. Kobychev, B.N. Kropivyansky, P.R. Maurenzig, L.L. Nagornaya, S.S. Nagorny, 
A.S. Nikolaiko, A.A. Pavlyuk, D.V. Poda, I.M. Solsky, M.V. Sopinskyy, Yu.G. Stenin, F. Taccetti, 
V.I. Tretyak, Ya.V. Vasiliev, S.S. Yurchenko, 
\textit{``Further study of CdWO$_4$ crystal scintillators as detectors for high sensitivity $2\beta$ 
experiments: Scintillation properties and pulse-shape discrimination''}, 
Nucl. Instr. Meth. A 569 (2006) 743${-}$753.

\vskip 0.4cm
\noindent [8] F.A. Danevich, V.V. Kobychev, S.S. Nagorny, D.V. Poda, V.I. Tretyak, S.S. Yurchenko, 
Yu.G. Zdesenko, 
\textit{``ZnWO$_4$ crystals as detector for $2\beta$ decay and dark matter experiments''}, 
Nucl. Instr. Meth. A 544 (2005) 553${-}$564.

\vskip 0.4cm
\noindent [9] I.I. Veretyannikov, F.A. Danevich, Yu.G. Zdesenko, V.V. Kobychev, S.S. Nagorny, D.V. Poda, 
\textit{``CdWO$_4$ scintillation detector optimization for the $2\beta$ experiment with $^{116}$Cd''} [in Ukrainian], 
Scientific Papers of KINR 13 (2004) 163${-}$172.

\clearpage
\newpage

\begin{center}
\textbf{CONTENTS}
\end{center}

\vskip 0.4cm
\noindent \textbf{List of symbols}\dotfill\ 6

\vskip 0.3cm
\noindent \textbf{Introduction}\dotfill\ 10

\vskip 0.2cm
\noindent \hspace{0.5cm} Relevance of the topic\dotfill\ 10

\noindent \hspace{0.5cm} Relation with academic programs, plans, and themes\dotfill\ 12

\noindent \hspace{0.5cm} The purpose and objectives of the study\dotfill\ 13

\noindent \hspace{0.5cm} Methods\dotfill\ 14

\noindent \hspace{0.5cm} Reliability\dotfill\ 15

\noindent \hspace{0.5cm} The novelty of the results\dotfill\ 16

\noindent \hspace{0.5cm} The scientific and practical significance of the obtained results\dotfill\ 17

\noindent \hspace{0.5cm} Personal contribution\dotfill\ 18

\noindent \hspace{0.5cm} Approbation of the results\dotfill\ 19

\noindent \hspace{0.5cm} Publications\dotfill\ 21

\vskip 0.3cm
\noindent \textbf{Section 1. Neutrino properties and double beta decay of atomic nuclei}\dotfill\ 22

\vskip 0.2cm
\noindent \hspace{0.5cm} 1.1. Neutrino mass and oscillation\dotfill\ 22 

\noindent \hspace{1cm} 1.1.1. Results of neutrino oscillation experiments\dotfill\ 23

\noindent \hspace{1cm} 1.1.2. The neutrino mass schemes\dotfill\ 25

\vskip 0.1cm
\noindent \hspace{0.5cm} 1.2. Double beta decay\dotfill\ 27

\noindent \hspace{1cm} 1.2.1. Various channels of $2\beta$ decay\dotfill\ 29

\noindent \hspace{1cm} 1.2.2. The half-lives of $2\beta$ processes\dotfill\ 30

\noindent \hspace{1cm} 1.2.3. Calculations of the nuclear matrix elements\dotfill\ 32

\vskip 0.1cm
\noindent \hspace{0.5cm} 1.3. Experimental methods of $2\beta$ decay study\dotfill\ 33

\noindent \hspace{1cm} 1.3.1. Requirements for detectors\dotfill\ 33

\noindent \hspace{1cm} 1.3.2. Geochemical and radiochemical experiments\dotfill\ 35

\noindent \hspace{1cm} 1.3.3. Counting experiments and their classification\dotfill\ 38

\vskip 0.1cm
\noindent \hspace{0.5cm} 1.4. The most sensitive $2\beta$ experiments\dotfill\ 39

\noindent \hspace{1cm} 1.4.1. Study of $2\beta^-$ decay\dotfill\ 39

\noindent \hspace{1.8cm} 1.4.1.1. Searches for $0\nu 2\beta^-$ decay of $^{76}$Ge in the IGEX and 
the Heidelberg- \hfill

\noindent \hspace{3.3cm} Moscow experiments with HPGe detectors\dotfill\ 42

\noindent \hspace{1.8cm} 1.4.1.2. The NEMO-3 calorimeter\dotfill\ 45

\noindent \hspace{1.8cm} 1.4.1.3. The Solotvina experiment with CdWO$_4$ scintillators\dotfill\ 46

\noindent \hspace{1.8cm} 1.4.1.4. Detector CUORICINO with TeO$_2$ crystals\dotfill\ 49

\noindent \hspace{1cm} 1.4.2. Searches for $2\varepsilon$ capture, $\varepsilon \beta^+$ and 
$2\beta^+$ decays\dotfill\ 50

\noindent \hspace{1.8cm} 1.4.2.1. Search for $2\varepsilon$ capture in $^{40}$Ca by using CaF$_2$(Eu) 
scintillators\dotfill\ 52

\noindent \hspace{1.8cm} 1.4.2.2. Searches for $\varepsilon \beta^+$ decay of $^{64}$Zn isotope\dotfill\ 53

\noindent \hspace{1.8cm} 1.4.2.3. The TGV-II experiment to search for $2\nu 2\varepsilon$ capture 
in $^{106}$Cd\dotfill\ 55

\noindent \hspace{1cm} 1.4.3. Study of $2\beta$ transitions to excited levels of daughter nuclei\dotfill\ 55

\vskip 0.1cm
\noindent \hspace{0.5cm} 1.5. Conclusions of section\dotfill\ 59

\vskip 0.3cm
\noindent \textbf{Section 2. Development of scintillation method to search for $0\nu 2\beta^-$ decay 
of}\hfill

\noindent \textbf{$^{116}$Cd isotope}\dotfill\ 61

\vskip 0.2cm
\noindent \hspace{0.5cm} 2.1. Development of low background scintillation detector with high energy 
resolution\dotfill\ 61

\newpage
\noindent \hspace{1cm} 2.1.1. Influence of the energy resolution on sensitivity of the experiment to search \hfill

\noindent \hspace{2.2cm} for $0\nu 2\beta^-$ decay of $^{116}$Cd on \dotfill\ 61
 
\noindent \hspace{1cm} 2.1.2. Analysis of the required energy resolution of CdWO$_4$ 
scintillation detector to \hfill

\noindent \hspace{2.2cm} search for $0\nu 2\beta^-$ decay of $^{116}$Cd\dotfill\ 63

\noindent \hspace{1cm} 2.1.3. Study of light collection and energy resolution of CdWO$_4$  
detector\dotfill\ 67

\noindent \hspace{1.8cm} 2.1.3.1. Experimental technique of measurements\dotfill\ 67

\noindent \hspace{1.8cm} 2.1.3.2. Measurements of light collection and energy resolution for 
different \hfill

\noindent \hspace{3.3cm} geometries of CdWO$_4$ detector\dotfill\ 69

\noindent \hspace{1.8cm} 2.1.3.3. Calculation of light collection for various geometries 
of detector\dotfill\ 71

\vskip 0.1cm
\noindent \hspace{0.5cm} 2.2. Analysis of PbWO$_4$ scintillators' applicability in 
experiment to search for $0\nu 2\beta^-$ \hfill

\noindent \hspace{1.4cm} decay of $^{116}$Cd\dotfill\ 73

\noindent \hspace{1cm} 2.2.1. Study of possibility to use PbWO$_4$ crystal as a light-guide for 
CdWO$_4$ \hfill

\noindent \hspace{2.2cm} scintillator\dotfill\ 73

\noindent \hspace{1cm} 2.2.2. Simulation of background of CdWO$_4$ detector surrounded by 
PbWO$_4$ \hfill

\noindent \hspace{2.2cm} scintillators\dotfill\ 76

\noindent \hspace{1cm} 2.2.3. Application of PbWO$_4$ scintillators in experiment 
to search for $0\nu 2\beta^-$ \hfill

\noindent \hspace{2.2cm} decay of $^{116}$Cd\dotfill\ 79

\vskip 0.1cm
\noindent \hspace{0.5cm} 2.3. Conclusions of section\dotfill\ 83

\vskip 0.3cm
\noindent \textbf{Section 3. Investigation of scintillation properties and radiopurity of the} \hfill

\noindent \textbf{detectors based on cadmium, lead, and zinc tungstate crystal scintillators}\dotfill\ 85

\vskip 0.2cm
\noindent \hspace{0.5cm} 3.1. Study of scintillation properties and radiopurity of  
CdWO$_4$ scintillators\dotfill\ 87

\noindent \hspace{1cm} 3.1.1. Measurements of the energy resolution with different CdWO$_4$ samples\dotfill\ 87

\noindent \hspace{1cm} 3.1.2. Scintillation response at low energies\dotfill\ 89

\noindent \hspace{1cm} 3.1.3. Measurements of $\alpha / \beta$ ratio for CdWO$_4$ scintillator\dotfill\ 91

\noindent \hspace{1cm} 3.1.4. Study of scintillation decay time\dotfill\ 93

\noindent \hspace{1cm} 3.1.5. Pulse shape discrimination of $\gamma$ quanta and 
$\alpha$ particles\dotfill\ 95

\noindent \hspace{1cm} 3.1.6. Dependence of pulse shape on temperature\dotfill\ 98

\noindent \hspace{1cm} 3.1.7. Study of radioactive contamination of CdWO$_4$ scintillators\dotfill\ 100

\noindent \hspace{1.8cm} 3.1.7.1. The time-amplitude analysis\dotfill\ 101

\noindent \hspace{1.8cm} 3.1.7.2. The pulse shape discrimination method\dotfill\ 103

\noindent \hspace{1.8cm} 3.1.7.3. Simulation of background\dotfill\ 104

\vskip 0.1cm
\noindent \hspace{0.5cm} 3.2. Investigation of scintillation properties and radioactive contamination
of PbWO$_4$ \hfill

\noindent \hspace{1.4cm} scintillators\dotfill\ 106

\noindent \hspace{1cm} 3.2.1. Doping PbWO$_4$ scintillators with different activators\dotfill\ 106

\noindent \hspace{1cm} 3.2.2. Dependence of light output and the energy resolution of 
PbWO$_4$\hfill

\noindent \hspace{2.2cm} crystals on dopants\dotfill\ 107

\noindent \hspace{1cm} 3.2.3. Measurement $\alpha / \beta$ ratio for doped PbWO$_4$ samples\dotfill\ 110

\noindent \hspace{1cm} 3.2.4. Pulse shape of PbWO$_4$ scintillators depending on the composition 
and \hfill

\noindent \hspace{2.2cm} concentration of dopants\dotfill\ 113

\noindent \hspace{1cm} 3.2.5. Pulse shape discrimination with doped PbWO$_4$ scintillators\dotfill\ 116

\noindent \hspace{1cm} 3.2.6. Measurements of radioactive contamination of PbWO$_4$ crystal 
scintillators\dotfill\ 119

\vskip 0.1cm
\noindent \hspace{0.5cm} 3.3. Study of scintillation properties and radiopurity of 
ZnWO$_4$ crystal scintillators \dotfill\ 121

\noindent \hspace{1cm} 3.3.1. Energy resolution and relative light output of ZnWO$_4$ 
scintillators\dotfill\ 121

\noindent \hspace{1cm} 3.3.2. Response of ZnWO$_4$ scintillator to $\gamma$ quanta and 
$\alpha$ particles\dotfill\ 123

\noindent \hspace{1cm} 3.3.3. Decay time of ZnWO$_4$ scintillator and pulse shape discrimination\dotfill\ 125

\noindent \hspace{1cm} 3.3.4. Development of ZnWO$_4$ crystals for low-background experiments\dotfill\ 126

\noindent \hspace{1.8cm} 3.3.4.1. Influence of dopants on scintillation properties of ZnWO$_4$ 
scintillators\dotfill\ 126

\noindent \hspace{1.8cm} 3.3.4.2. Investigation of radiopurity of ZnWO$_4$ crystals\dotfill\ 128

\vskip 0.1cm
\noindent \hspace{0.5cm} 3.4. Conclusions of section\dotfill\ 132

\vskip 0.3cm
\noindent \textbf{Section 4. Search for $2\beta$ decay of Zinc and Tungsten isotopes with the help}\hfill

\noindent \textbf{of ZnWO$_4$ scintillators}\dotfill\ 133

\vskip 0.2cm
\noindent \hspace{0.5cm} 4.1. Experimental studies with ZnWO$_4$ crystal scintillators\dotfill\ 133

\noindent \hspace{1cm} 4.1.1. Low-background ``DAMA/R\&D'' set-up\dotfill\ 133

\noindent \hspace{1cm} 4.1.2. Stages of low-background measurements with ZnWO$_4$ scintillators\dotfill\ 135

\noindent \hspace{1cm} 4.1.3. The ICP-MS analysis of contaminations of ZnWO$_4$ crystals\dotfill\ 137

\vskip 0.1cm
\noindent \hspace{0.5cm} 4.2. Analysis of background of ZnWO$_4$ scintillators\dotfill\ 139

\noindent \hspace{1cm} 4.2.1. PMT noise rejection\dotfill\ 140

\noindent \hspace{1cm} 4.2.2. Determination of radionuclides' activity by the time-amplitude analysis\dotfill\ 141

\noindent \hspace{1cm} 4.2.3. Pulse shape discrimination of $\gamma$ quanta and 
$\alpha$ particles\dotfill\ 142

\noindent \hspace{1cm} 4.2.4. Simulation of background of ZnWO$_4$ scintillators\dotfill\ 144

\vskip 0.1cm
\noindent \hspace{0.5cm} 4.3. Investigation of $2\beta$ processes in Zn and W isotopes\dotfill\ 147

\noindent \hspace{1cm} 4.3.1. Search for $\varepsilon \beta^+$  decay of $^{64}$Zn isotope\dotfill\ 150

\noindent \hspace{1cm} 4.3.2. Double electron capture in $^{64}$Zn and $^{180}$W isotopes\dotfill\ 152

\noindent \hspace{1cm} 4.3.3. Study of $2\beta^-$ decay of $^{70}$Zn and $^{186}$W isotopes\dotfill\ 154

\noindent \hspace{1cm} 4.3.4. Comparison of the obtained half-life limits with the results of 
previous \hfill

\noindent \hspace{2.2cm} experiments and theoretical predictions\dotfill\ 157

\vskip 0.1cm
\noindent \hspace{0.5cm} 4.4. Conclusions of section\dotfill\ 161

\vskip 0.3cm
\noindent \textbf{Conclusions}\dotfill\ 163

\vskip 0.3cm
\noindent \textbf{Acknowledgements}\dotfill\ 166

\vskip 0.3cm
\noindent \textbf{References}\dotfill\ 167

\end{document}